\documentclass[10pt,aps,prd,showpacs,nofootinbib,groupedaddress]{revtex4-1}
\usepackage{hyperref}
\usepackage{amsfonts}
\usepackage{amsmath}
\usepackage{amssymb}
\usepackage{graphicx}
\usepackage{color}
\usepackage{textcomp}

\def\be{\begin{equation}}
\def\te{\end{equation}}
\def\ee{\end{equation}}
\def\ba{\begin{eqnarray}}
\def\bea{\begin{eqnarray}}
\def\nn{\nonumber\\}
\def\tea{\end{eqnarray}}
\def\ea{\end{eqnarray}}
\def\eea{\end{eqnarray}}

\usepackage{amssymb}

\begin{document}

\title{The importance of being measurement}

\author{Esteban Calzetta}
\email{calzetta@df.uba.ar}
\affiliation{Departamento de F\'isica, Facultad de Ciencias Exactas y Naturales, Universidad de Buenos Aires and IFIBA, 
CONICET, Cuidad Universitaria, Buenos Aires 1428, Argentina}


\begin{abstract}
Energy exchanges under form of heat is neither the most natural or efficient way to operate an engine in the quantum realm. Recently there have been in the literature several proposals for ``quantum measurement engines'' where energy is fed into the machine by operations which otherwise would be conducive to quantum measurements on the working substance (henceforth ``the system''). In the analysis of the working of these devices, oftentimes it is assumed that the only effect of measurement is to turn the state of the system from whatever prior state to an eigenstate of the measured property, and energy exchanges are determined therefrom. This ignores the intricacies of the quantum measurement process. We propose a simple model of a quantum measurement engine where the measurement process may be analyzed in detail, and therefore energy exchanges, and limitations on their duration, may be traced more fully. 
\end{abstract}

\maketitle
\section{Introduction}
The development of engines working at the nano scale is one of the most fascinating challenges facing our discipline \cite{Lutz12,Lutz14,Christian}. While it is natural to draw on our substantial knowledge of macroscopic machines as a guide to understanding, the fact is that building replicas of large scale machines is not necessarily the best strategy in the quantum domain. Along this line, it has been suggested that ``thermal'' engines, where energy is fed into the machine under heat form, with the concomitant limitations associated to the Second Law and the speed limits for heat exchange, may be replaced (and outperformed) by ``quantum measurement'' engines, where the coupling to the machine is performed by devices originally meant to carry out measurements on the system \cite{Talkner17,Talkner18,Das18,Kosloff19}.

To give an accurate analysis of the working of a quantum measurement engine, and particularly to discuss their limitations, if any, it is essential to take into account the intricacies of quantum measurement apparatuses \cite{Mittelstaedt98,Wiseman09,Jacobs14,Busch16,Balian13,Pasquale19}. A quantum measurement device is a rather complicated thing, maybe coupled to its own environment \cite{Eli,JP}, and which certainly has the means to build and hold a record of what has been measured \cite{Hartle93}. Therefore, quantum measurement involves energy exchanges other than to and from the system, and not all the exchanged energy may be retrievable after the process is completed. Those energy exchanges must be considered  in the analysis, as well as whether there are limitations on the time necessary for their completion.

As a case in point, we propose in this note a simple model of a quantum measurement engine where the measurement process may be analyzed in detail. The machine is a spin one half particle operating under a quantum Otto cycle \cite{Rezek17,Erdman18,Ankerhold19}. In the quantum measurement version, the contact with the hot reservoir in the Otto cycle is replaced by a quantum measurement of the spin. This is not a ``Maxwell Demon'' type engine \cite{Anders17a,Anders17b,Elouard17,Elouard18,Kim18,Seah19}, the measurement process is carried out only to feed energy into the machine but there is no feedback from the measurement outcome.

The measurement is carried out by making the spin precede around an auxiliary magnetic field, thus emitting electromagnetic radiation. The state of the spin is recorded into the state of the outgoing radiation. In this preliminary study we shall not consider the back reaction of the radiation process on the spin, but only the energy carried out by the radiation, and the time limitations on the process if a successfull measurement is assumed.

As a matter of fact, a precise measurement, namely, that the two possible initial states of the spin lead to mutually orthogonal states of the radiation, requires either an infinite energy output or an infinite measurement time. This is consistent with formal analysis of the quantum measurement process \cite{Guryanova18}.

Given that a compromise shall be reached, we analysis the impact of the energy cost of measurement on the machine efficiency and its power-efficiency relationship.

This note is organized as follows. In the next section we describe the quantum Otto cycle which is our standard heat engine. In Section III we turn this into a quantum measurement engine by replacing contact with a hot reservoir in the Otto cycle by a spin measurement. Section IV is the core of the note because here we analyze in detail the measurement process and its cost in energy and in time. We conclude with some very brief final remarks.

\section{A quantum Otto cycle}
Let us begin by describing the thermal engine which will serve as a contrast to the quantum measurement engine to be introduced below. This will be a simple implementation of a quantum Otto cycle.

The system is a spin $1/2$ particle. At the beginning of the cycle it is in a thermal state at temperature $T$ coupled to a magnetic field $B_z\left( 0\right) $ in the $z$ direction. We understand this to mean that the spin in the $z$ direction is well defined, and takes the value  $1/2$ with probability

\be
p_+=\frac1{1+e^{-\beta_0\mu B_z\left( 0\right) }}>\frac12
\te 
and the value $-1/2$ with probability $p_-=1-p_+<p_+$. Here $\beta =1/k_BT$, $k_B$ is Boltzmann constant, and 

\be 
\mu=\frac{e\hbar}{2m}
\te 
where $e$ and $m$ are the charge and mass of the particle. A gyromagnetic factor $g=2$ is assumed. Also $e^2=\alpha\hbar c$, where $\alpha =1/137$. With these choices magnetic field has units of $QTL^{-3}$. The mean energy is 

\be 
\left\langle E\right\rangle \left( 0\right) =-\mu B_z\left( 0\right)\left( p_+-p_-\right)  
\te 
In the first leg of the cycle, we increase the field adiabatically to a value $B_z\left( 1\right) $. The entropy does not change, and the mean energy decreases to 

\be 
\left\langle E\right\rangle \left( 1\right) =-\mu B_z\left( 1\right)\left( p_+-p_-\right)  
\te 
therefore work is obtained from the machine, at a value

\be 
W_{01}=-\left(\left\langle E\right\rangle \left( 1\right)-\left\langle E\right\rangle \left( 0\right)\right) =\mu\left(B_z\left( 1\right)-B_z\left( 0\right)\right) \left( p_+-p_-\right) 
\te 
In the second leg, we bring the spin to a state  where the spin in the $z$ direction is well defined and takes either value with probability $1/2$. This could be achieved by coupling the system to a heath bath at infinite temperature. In any case, the heat exchange is irreversible, because the system is not at the bath temperature. The new mean energy is 

\be 
\left\langle E\right\rangle \left( 2\right) =0  
\te 
so the heath exchanged is 

\be
Q_{12}=\left\langle E\right\rangle \left( 2\right)-\left\langle E\right\rangle \left( 1\right)=-\left\langle E\right\rangle \left( 1\right)=\mu B_z\left( 1\right)\left( p_+-p_-\right)  
\te 
In the third leg, we bring the field adiabatically back to $B_z\left( 0\right) $. Since $p_+=p_-$ throughout, no net work is exchanged. Finally, we allow the system to thermalize again, emitting a heat 

\be 
Q_{30}=-\left\langle E\right\rangle \left( 0\right)
\te
Obviously

\be 
W_{01}+Q_{30}=Q_{12}
\te 
and we may define an efficiency

\be 
\eta_O = \frac{W_{01}}{Q_{12}}=1-\frac{\left\langle E\right\rangle \left( 0\right)}{\left\langle E\right\rangle \left( 1\right)}=1-b
\te
where

\be
b=\frac{B_z\left( 0\right) }{B_z\left( 1\right) }
\te
It is difficult to give an estimate of the power limitations on the machine, given the possibility to recur to shortcuts on all and any of the four legs in the cycle \cite{Muga19,yo,Cakmak19,Villazon19,Adolfo19}. To obtain a simple estimate we may assume that heat exchanges may be made instantaneous. On the other hand, the duration of the $0-1$ and $2-3$ legs is restricted by the adiabaticity condition

\be 
\frac{\dot E}E\ll \omega\approx \frac E{\hbar}
\te 
or 

\be 
\frac d{dt}\frac1E\ll\frac 1{\hbar}
\te 
leading to 

\be 
\Delta t\gg \frac{\hbar}{\mu}\left( \frac1{B_z\left( 0\right) } - \frac1{B_z\left( 1\right)}\right) 
\te 
While a conventional Quantum Speed Limit estimate \cite{Mandelstam45,Lloyd03} yields 

\be 
\Delta t\ge\frac{\hbar}{\mu\left( {B_z\left( 1\right) } -{B_z\left( 0\right)}\right) }
\te 
We therefore estimate

\be
\mathrm{Power}\;=P_{max}\left[\frac{b\left(1-b\right)^2}{b+\left(1-b\right)^2}\right]
\te
where

\be
P_{max}=\frac{\left(\mu B_z\left(1\right)\right)^2\left(p_+-p_-\right)}{\hbar}
\te
We obtain maximum power $\approx 0.19P_{max}$ for $b\approx 0.36$

\section{Quantum measurement implementation}
We wish to replace the step $1\to 2$ by an interaction. The obvious choice is to polarize the spin in the $y$ direction, by applying a strong field in that direction. Once the $y$ field is removed, the spin is left pointing in the $y$ direction, and indeed both $z$ projections occur with the same probability. Let us see whether it works.

We consider a Pauli spinor evolving under the Hamiltonian

\be
H=-g\mu \left[ B_z\sigma_z+B_y\left(t\right)\sigma_y\right] 
\te
with instantaneous eigenvalues $\pm\hbar\omega$, $\omega=\mu \sqrt{B_z^2+B_y^2\left(t\right)}/\hbar$. The instantaneous eigenstates have well defined spin along the direction $\hat v=\cos\theta \hat z+\sin\theta\hat y$, where $\cos\theta=B_z/\omega$. This direction is obtained by a rotation of angle $-\theta$ around the $\hat x$ direction, thus the instantaneous eigenvectors are 

\be 
\phi_+=\left(\begin{array}{c}\cos\frac12\theta\\i\sin\frac12\theta\end{array}\right)
\te
and 

\be 
\phi_-=\left(\begin{array}{c}i\sin\frac12\theta\\\cos\frac12\theta\end{array}\right)
\te
They obey $\dot\phi_+=i\dot\theta\phi_-/2$ and $\dot\phi_-=i\dot\theta\phi_+/2$. 
For example, consider the case when $B_y$ is suddenly turned on at $t=0$ and then turned off at $t_0$. If the state at $t=0^-$ is $s_+=\left(1,0\right)$, then for $t\ge 0$ we have

\bea
\Psi_+\left(t\right)&=&\cos\frac12\theta e^{i\omega t}\phi_+-i\sin\frac12\theta e^{-i\omega t}\phi_-\nn
&=&\psi_+\left(t\right)s_+-\psi_-\left(t\right)s_-
\tea
where

\bea
\psi_+\left(t\right)&=& \cos\omega t +i\sin\omega t \cos\theta\nn
\psi_-\left(t\right)&=&\sin\omega t\sin\theta 
\tea
Then  from $t_0^+$ on 

\be
\Psi_+\left(t\right)=\psi_+\left(t_0\right)e^{i\mathcal B}s_+-\psi_-\left(t_0\right)e^{-i\mathcal B}s_-
\te
where $\mathcal B\left(t_0\right)=0$ and

\be
\dot{\mathcal B}=\frac{\mu B_z}{\hbar}
\te
Therefore

\be
\left\langle E\right\rangle_+ =-\mu B_z\left(1\right)\left[1-2\sin^2\theta\sin^2\omega t_0\right]
\te 
If the initial state is $s_-$ then at $t\ge 0$

\bea
\Psi_-\left(t\right)&=&-i\sin\frac12\theta e^{i\omega t}\phi_++\cos\frac12\theta e^{-i\omega t}\phi_-\nn
&=&\psi_-\left(t\right) s_++\psi^*_+\left(t\right)s_-
\tea
and for $t\ge t_0$

\be
\Psi_-\left(t\right)=\psi_-\left(t_0\right)e^{i\mathcal B}s_++\psi^*_+\left(t_0\right)e^{-i\mathcal B}s_-
\te
so 

\be
\left\langle E\right\rangle_- =-\left\langle E\right\rangle_+
\te 
In this implementation we get

\be
\left\langle E\right\rangle\left(2\right)=-\mu B_z\left(1\right)\left(p_+-p_-\right)\left[1-2\sin^2\theta\sin^2\omega t_0\right]
\te

\be
Q_{12}=2\mu B_z\left(1\right)\left(p_+-p_-\right)\sin^2\theta\sin^2\omega t_0
\te
Now getting from $2$ to $3$ requires to do work on the system

\be
W_{23}=\mu \left(B_z\left(1\right)-B_z\left(0\right)\right)\left(p_+-p_-\right)\left[1-2\sin^2\theta\sin^2\omega t_0\right]
\te
On thermalization, the system sheds heat

\be
Q_{30}=2\mu B_z\left(0\right)\left(p_+-p_-\right)\sin^2\theta\sin^2\omega t_0
\te
Obviously

\be
W_{01}+Q_{30}=Q_{12}+W_{23}=\mu \left(p_+-p_-\right)\left[B_z\left(1\right)-B_z\left(0\right)\left(1-2\sin^2\theta\sin^2\omega t_0\right)\right]
\te
and the efficiency is

\be
\eta_Q=\frac{1-b}{1-b\left(1-2\sin^2\theta\sin^2\omega t_0\right)}
\te
which for $\omega t_0\gg 1$ may be approximated as

\be
\eta_Q=\frac{1-b}{1-b\cos^2\theta}
\te
It is clear that the quantum measurement engine has a definite potential for outperforming the thermal Otto cycle. We now want to validate this analysis by a more carefull consideration of the $1\to 2$ (measurement) leg.

\section{Focus on measurement}

The most important feature of the quantum measurement process is that it leaves a record of what has been measured \cite{Hartle93}. In our case we choose as recording device the electromagnetic radiation emitted by the time-varying spin. To obtain the quantum state of the radiation we shall proceed in two steps. First we will work out the expectation value of the radiation field for a given intial state of the spin, $s_+$ or $s_-$. Then we shall apply an appropriate displacement operator to the electromagnetic vacuum to match that expectation value. 

Let us first say something about the spin evolution. From $t=0$ to $t_0$ the $B_y$ field is turned on and the spin precedes around the $\left(B_y,B_z\left(1\right) \right)$ direction. From then on it precedes around the $z$ axis at a much lower rate, provided $B_y\gg B_z\left(1\right)\ge B_z\left(0\right)$. We shall neglect any radiation from this second leg. 

Let us consider a basis $\hat x'=\hat x$, $\hat y'=\cos\theta\hat y-\sin\theta\hat z$, $\hat z'=\sin\theta\hat y+\cos\theta\hat z$. Then $\left\langle S_{z'}\right\rangle_{\pm}=\pm \cos\theta/2$ is constant and does not contribute to the radiation field. Else, if the initial state is $s_+$, then

\bea
\left\langle S_{x'}\right\rangle_+&=&\frac{-1}2\sin\theta\sin 2\omega t\nn
\left\langle S_{y'}\right\rangle_+&=&\frac{1}2\sin\theta\cos 2\omega t
\tea
If the initial state is $s_-$, then $\left\langle S_i\right\rangle_-=-\left\langle S_i\right\rangle_+$. It follows that

\be
\left\langle \dot{\vec S}\right\rangle =2\omega\; \hat z'\times \left\langle{\vec S}\right\rangle 
\te
and

\be
\left\langle \ddot{\vec S}\right\rangle =-4\omega^2 \left\langle{\vec S}\right\rangle_{\perp} 
\te
where $\left\langle{\vec S}\right\rangle_{\perp}=\left\langle S_{x'}\right\rangle\hat x'+\left\langle S_{y'}\right\rangle\hat y'$. This is the part of the spin that radiates.

We now turn to the electromagnetic field. As we said before, we first consider tits expectation value, for a given initial value of the spin. Since electromagnetism is a linear   theory, the expectation value of the  vector potential $\vec A$ follows Maxwell equations sourced by the mean value of the magnetization. Choosing a gauge with $\nabla\vec A=0$, 

\be 
\frac1{c^2}\frac{\partial^2}{\partial t^2}\vec A-\mathbf{\Delta}\vec A=\frac{4\pi}{c^2}\vec j
\te
where $\vec j=\nabla\times\vec M$, and $\vec M$ is the magnetization density. We only consider the time-dependent part

\be
\vec M=g\mu\left\langle \vec S\left( t\right) \right\rangle_{\perp} \delta\left( \vec x\right) 
\te
So 

\be 
\vec A\left( \vec x,t\right) =-\nabla\times\frac{g\mu}{c^2r}\left\langle \vec S\left( t- \frac rc\right) \right\rangle_{\perp}=  \frac{g\mu}{c^3r^2}\vec r\times \left\langle \dot{\vec S}\left( t- \frac rc\right) \right\rangle_{\perp} +O\left(r^{-2}\right)
\te
We shall only consider the leading terms, corresponding to the radiation field. The electric field

\be
\vec E=-\dot{\vec A}= -\frac{g\mu}{c^3r^2}\vec r\times \left\langle \ddot{\vec S}\left( t- \frac rc\right) \right\rangle_{\perp} 
\te
The magnetic field

\be
\vec B=\nabla\times\vec A=\frac{g\mu}{c^3r}\left[\left\langle \ddot{\vec S}\left( t- \frac rc\right) \right\rangle_{\perp} -\frac1{r^2}\vec r\left(\vec r\cdot\left\langle \ddot{\vec S}\left( t- \frac rc\right) \right\rangle_{\perp} \right)\right]
\te
and the Poynting vector

\be
\vec{\mathbf{S}}=c\vec E\times\vec B=\frac{g^2\mu^2}{c^5r^3}\vec r\left[\left\langle \ddot{\vec S}\left( t- \frac rc\right)\right\rangle_{\perp}^2 -\frac1{r^2}\left(\vec r\cdot\left\langle \ddot{\vec S}\left( t- \frac rc\right) \right\rangle_{\perp}\right)^2\right]
\te
The power radiated at time $t$ through an sphere of radius $r$ is

\be
P=\frac{8\pi}3\frac{g^2\mu^2}{c^5}\left\langle \ddot{\vec S}\left( t- \frac rc\right)\right\rangle_{\perp}^2 
\te
This concludes that analysis of the expectation values. We shall now reconstruct the full quantum state of the radiation field. We assume the initial spin state is $s_+$, and write

\be
\left\langle S_{x'}\right\rangle\left(t\right)=\int\frac{df}{2\pi}\;e^{-ift}\;S_{x'}\left(f\right)
\te 
where

\be
S_{x'}\left(f\right)=\frac{-1}2\sin\theta\int_0^{t_0}dt\;e^{ift}\sin 2\omega t=\frac{\sin\theta}4\left[\frac{e^{i\left(f+2\omega\right)t_0}-1}{f+2\omega}-\frac{e^{i\left(f-2\omega\right)t_0}-1}{f-2\omega}\right]
\te
Similarly

\bea
\left\langle S_{y'}\right\rangle\left(t\right)&=&\int\frac{df}{2\pi}\;e^{-ift}\;S_{y'}\left(f\right)\nn
S_{y'}\left(f\right)&=&\frac{1}2\sin\theta\int_0^{t_0}dt\;e^{ift}\cos 2\omega t=\frac{-i\sin\theta}4\left[\frac{e^{i\left(f+2\omega\right)t_0}-1}{f+2\omega}+\frac{e^{i\left(f-2\omega\right)t_0}-1}{f-2\omega}\right]
\tea 
Observe that $S_i\left(-f\right)=S_i\left(f\right)^*$. These expressions cannot be used at very high frequency, where the Fourier amplitudes depend on the way the $B_y$ field is turned on. We shall assume the Fourier amplitudes peak around $f\approx\omega$, where $S_{x'}\approx S_{y'}\approx t_0\sin\theta$. So

\be
\left\langle{\vec S}\right\rangle_{\perp}\delta\left(\vec x\right)=\int\frac{d^3k}{\left(2\pi\right)^3}\frac{df}{2\pi}\;e^{i\left(\vec k\vec x-ft\right)}\left[S_{x'}\left(f\right)\hat x'+S_{y'}\left(f\right)\hat y'\right]
\te
and the induced current is 

\be
\vec j=ig\mu\int\frac{d^3k}{\left(2\pi\right)^3}\frac{df}{2\pi}\;e^{i\left(\vec k\vec x-ft\right)}\left[S_{x'}\left(f\right)\vec k\times\hat x'+S_{y'}\left(f\right)\vec k\times\hat y'\right]
\te
It is convenient to introduce a triad $\left(\hat k,\epsilon^{\left(1\right)}_{\vec k},\epsilon^{\left(2\right)}_{\vec k}\right)$, where

\be
\epsilon^{\left(1\right)}_{\vec k}=\frac{\vec k\times\hat x'}{\left|\vec k\times\hat x'\right|}=\frac{k_{z'}\hat y'-k_{y'}\hat z'}{\sqrt{k^2-k_{x'}^2}}
\te

\be
\epsilon^{\left(2\right)}_{\vec k}=\hat k\times\epsilon^{\left(1\right)}_{\vec k}=\frac{-\left(k_{y'}^2+k_{z'}^2\right)\hat x'+k_{x'}\left(k_{y'}\hat y'+k_{z'}\hat z'\right)}{k\sqrt{k^2-k_{x'}^2}}
\te

Then $\epsilon^{\left(1\right)}_{-\vec k}=-\epsilon^{\left(1\right)}_{\vec k}$, $\epsilon^{\left(2\right)}_{-\vec k}=\epsilon^{\left(2\right)}_{\vec k}$, 

\be 
\hat y'=\hat k\left( \hat k\cdot\hat y'\right) +\epsilon^{\left(1\right)}_{\vec k}\left(\epsilon^{\left(1\right)}_{\vec k} \cdot\hat y'\right)+\epsilon^{\left(2\right)}_{\vec k}
\left(\epsilon^{\left(2\right)}_{\vec k} \cdot\hat y'\right)=\frac{k_{y'}}k\hat k+\frac{k_{z'}}{\sqrt{k^2-k_{x'}^2}}\epsilon^{\left(1\right)}_{\vec k}+\frac{k_{x'}k_{y'}}{k\sqrt{k^2-k_{x'}^2}}\epsilon^{\left(2\right)}_{\vec k}
\te 
and 

\be 
\vec k\times\hat y'=\frac{kk_{z'}}{\sqrt{k^2-k_{x'}^2}}\epsilon^{\left(2\right)}_{\vec k}-\frac{k_{x'}k_{y'}}{\sqrt{k^2-k_{x'}^2}}\epsilon^{\left(1\right)}_{\vec k}
\te 
Finally we get the induced current as

\be
\vec j=ig\mu\sum_{\alpha}\int\frac{d^3k}{\left(2\pi\right)^3}\frac{df}{2\pi}\;e^{i\left(\vec k\vec x-ft\right)}S_{\alpha}\left(f,\vec k\right)\epsilon^{\left(\alpha\right)}_{\vec k}
\te
where 

\bea
S_{1}\left(f,\vec k\right)&=&\sqrt{k^2-k_{x'}^2}S_{x'}\left(f\right)-\frac{k_{x'}k_{y'}}{\sqrt{k^2-k_{x'}^2}}S_{y'}\left(f\right)\nn
S_{2}\left(f,\vec k\right)&=&\frac{kk_{z'}}{\sqrt{k^2-k_{x'}^2}}S_{y'}\left(f\right)
\tea
The solution to Maxwell equations reads

\be 
\vec A=-4\pi ig\mu\sum_{\alpha}\int\frac{d^3k}{\left(2\pi\right)^3}\frac{df}{2\pi}\;e^{i\left(\vec k\vec x-ft\right)}\frac{S_{\alpha}\left(f,\vec k\right)\epsilon^{\left(\alpha\right)}_{\vec k}}{\left( f+i\epsilon\right) ^2-c^2k^2}
\te
The radiation field is obtained by keeping only the contributions from the poles 

\be 
\vec A_{rad}=-4\pi g\mu\sum_{\alpha}\int\frac{d^3k}{\left(2\pi\right)^3}\;\frac{e^{i\vec k\vec x}}{2ck}\left[ e^{-ickt}S_{\alpha}\left(ck,\vec k\right)\epsilon^{\left(\alpha\right)}_{\vec k}+e^{ickt}S_{\alpha}^*\left(ck,-\vec k\right)\epsilon^{\left(\alpha\right)}_{-\vec k}\right] 
\te
Comparing this to the plane wave expansion of the free Heisenberg operator \cite{Bjorken}

\be 
\vec {\mathbf A}=\sqrt\hbar\sum_{\alpha}\int\frac{d^3k}{\left(2\pi\right)^3}\;\frac{e^{i\vec k\vec x}}{\sqrt{2ck}}\left[ e^{-ickt}a_{\alpha,\vec k}\epsilon^{\left(\alpha\right)}_{\vec k}+e^{ickt}a^{\dagger}_{\alpha,-\vec k}\epsilon^{\left(\alpha\right)}_{-\vec k}\right] 
\te
we see that the quantum state of the radiation field may be obtained by the displacement $a_{\alpha,\vec k}\to \left\langle a_{\alpha,\vec k}\right\rangle+a'_{\alpha,\vec k}$, where 

\be
\left\langle a_{\alpha,\vec k}\right\rangle=-\frac{4\pi g\mu}{\sqrt{2\hbar ck}}S_{\alpha}\left(ck,\vec k\right)
\te
The displaced state is
\be
\left|rad\right\rangle_+=e^{R}\left|0\right\rangle
\te
where

\be
R=4\pi g\mu\sum_{\alpha}\int\frac{d^3k}{\left(2\pi\right)^3}\;\frac1{\sqrt{2\hbar ck}}\left[ S^*_{\alpha}\left(ck,\vec k\right)a_{\alpha,\vec k}-S_{\alpha}\left(ck,\vec k\right)a^{\dagger}_{\alpha,\vec k}\right] 
\te
If the initial spin state is $s_-$, then the final radiation state is

\be
\left|rad\right\rangle_-=e^{-R}\left|0\right\rangle
\te
Observe that $_+\left\langle rad\vert rad\right\rangle_+=_-\left\langle rad\vert rad\right\rangle_-=1$ while

\be
_-\left\langle rad\vert rad\right\rangle_+=\left\langle 0\right|e^{2R}\left|0\right\rangle=e^{-\Gamma}
\te
where

\be
\Gamma=\frac{16\pi^2 g^2\mu^2}{\hbar c}\sum_{\alpha}\int\frac{d^3k}{\left(2\pi\right)^3}\;\frac1{k}S^*_{\alpha}\left(ck,\vec k\right)S_{\alpha}\left(ck,\vec k\right)
\te
Under the assumption that the Fourier amplitudes peak at frequencies of the order of $\omega$, the integral may be evaluated as $\approx \omega^4t_0^2\sin^2\theta /c^4$. Then the total radiated energy is $\approx\Gamma\hbar /t_0$.

\begin{figure}
\includegraphics[scale=0.5]{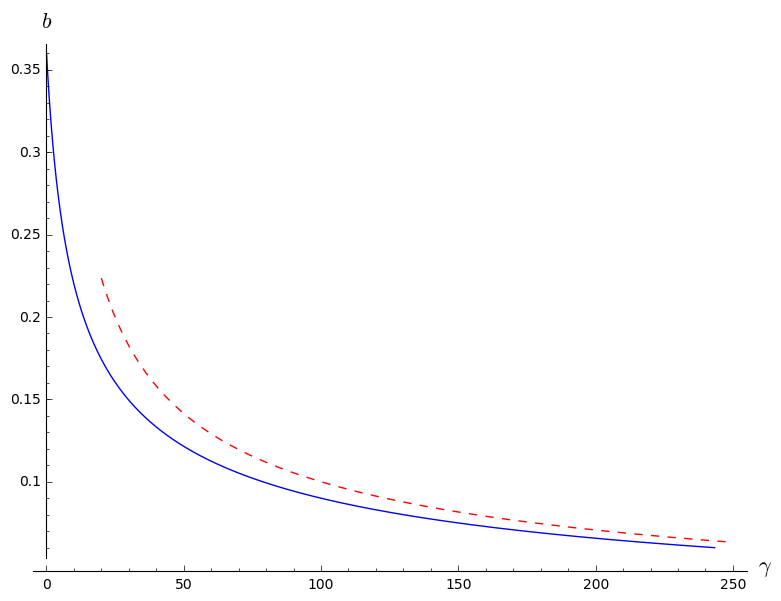}
\caption{[Color online] (full line) The value of $b$ for which we obtain maximum power, as a function of $\gamma$, defined in Eq. (\ref{gamma}); (dashes) for comparison, the line $1/\sqrt{\gamma}$. }
\label{bmax}
\end{figure}

\begin{figure}
\includegraphics[scale=0.5]{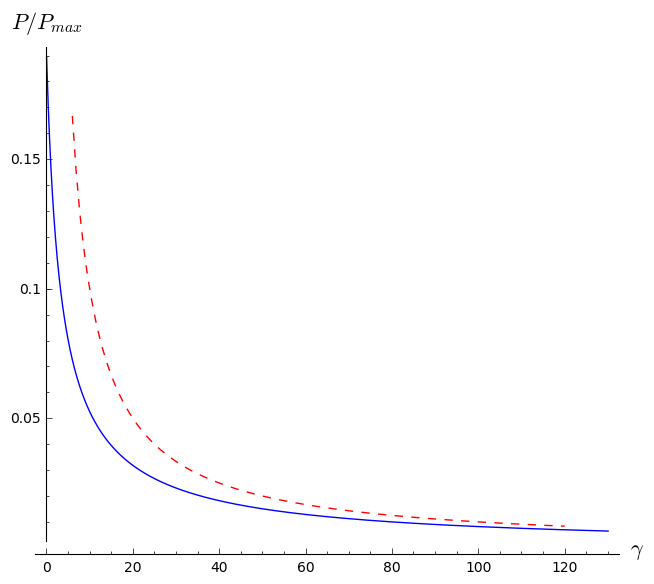}
\caption{[Color online] (full line) Maximum power as a fraction of $P_{max}$, as a function of $\gamma$, defined in Eq. (\ref{gamma}); (dashes) for comparison, the line $1/{\gamma}$. }
\label{Pmax}
\end{figure}

We see that a precise measurement of the spin requires $\Gamma\to\infty$ and thus infinite resources, as expected on formal grounds \cite{Guryanova18}. Otherwise, we obtain a more accurate estimate for the machine efficiency

\be
\eta'_Q=\frac{1-b}{1-b\cos^2\theta+\gamma^{-1}\Gamma}
\te
where

\be
\gamma=\frac{\mu B_z\left(1\right)t_0}{\hbar}
\label{gamma}
\te
The efficiency is reduced from the previous estimate but still may outperform the quantum Otto cycle. We must also correct our power estimate

\be
\mathrm{Power}'\;=P_{max}\left[\frac{b\left(1-b\right)^2}{b+\left(1-b\right)^2+\gamma b\left(1-b\right)}\right]
\te
The value of $b$ for which we obtain maximum power now depends on $\gamma$ (see Fig.(\ref{bmax})), and the maximum power is always less than $0.19P_{max}$ (see Fig.(\ref{Pmax}))

\section{Final remarks}
In this note we have presented a simple model of a quantum measurement engine whereby different energy exchanges may be tracked, over and above those from and to the working substance itself. We have shown that the energy necessary to build a record of the measurement result (in our example, the record being the quantum state of the radiation field) has a definite impact on the expected efficiency of the engine. There are also limitations on the time necessary to perform the measurement, which likewise affect the power which the engine may produce. Notwithstanding, the possibility of a quantum measurement engine outperforming a thermal one remains. Interestingly, a good engine requires a rather poor measurement, and viceversa.

Our analysis may be improved in many ways, most obviously in including the back reaction of radiation on the spin itself. We expect to proceed with these improvements in future work.

\acknowledgements 
Work supported in part by Universidad de Buenos Aires and CONICET (Argentina).

It is a pleasure to acknowledge exchanges with the QUFIBA and LIAF groups at the Physics Department, FCEN-UBA (Argentina)

\end{document}